\documentclass[twocolumn,showpacs,preprintnumbers,amsmath,amssymb]{revtex4}

\usepackage{graphicx}
\usepackage{dcolumn}
\usepackage{bm}

\begin{document}


\title {Phase structure of a surface model on dynamically triangulated spheres with elastic skeletons}

\author{Hiroshi Koibuchi}
 \email{koibuchi@mech.ibaraki-ct.ac.jp}
\affiliation{%
Department of Mechanical and Systems Engineering, Ibaraki National College of Technology, Nakane 866 Hitachinaka, Ibaraki 312-8508, Japan}%

\date{\today}

\begin{abstract}
We find three distinct phases; a tubular phase, a planar phase, and the spherical phase, in a triangulated fluid surface model. It is also found that these phases are separated by discontinuous transitions. The fluid surface model is investigated within the framework of the conventional curvature model by using the canonical Monte Carlo simulations with dynamical triangulations. The mechanical strength of the surface is given only by skeletons, and no two-dimensional bending energy is assumed in the Hamiltonian. The skeletons are composed of elastic linear-chains and rigid junctions and form a compartmentalized structure on the surface, and for this reason the vertices of triangles can diffuse freely only inside the compartments. As a consequence, an inhomogeneous structure is introduced in the model; the surface strength inside the compartments is different from the surface strength on the compartments. However, the rotational symmetry is not influenced by the elastic skeletons; there is no specific direction on the surface. In addition to the three phases mentioned above, a collapsed phase is expected to exist in the low bending rigidity regime that was not studied here. The inhomogeneous structure and the fluidity of vertices are considered to be the origin of such variety of phases. 
\end{abstract}

\maketitle
\section{Introduction}\label{intro}
A crumpling of surfaces has been investigated on the basis of the singularity analysis, and progress has been recently made on understanding the crumpling phenomena; the universal structure on the crumpled thin sheets was found in the formations of singularity of ridges and cones  \cite{CM-PRL1998,SCM-SCIE2000}. A similar transition to this phenomena was also found experimentally between the smooth state and the crumpled state in an artificial membrane, which is partly polymerized \cite{CNE-PRL-2006}. 

 Studies have also been focused on the transition in the surface model of Helfrich, Polyakov and Kleinert (HPK) \cite{HELFRICH-1973,POLYAKOV-NPB1986,KLEINERT-PLB1986} from the viewpoint of statistical  mechanics \cite{NELSON-SMMS2004-1,David-TDQGRS-1989,NELSON-SMMS2004-149,Wiese-PTCP2000,Bowick-PREP2001,Gompper-Schick-PTC-1994,WHEATER-JP1994}. The bending rigidity is known to be stiffened by the thermal fluctuation of the surface, and this was confirmed in the statistical mechanics of membranes \cite{Peliti-Leibler-PRL1985,DavidGuitter-EPL1988,PKN-PRL1988,BKS-PLA2000,BK-PRB2001}. Numerical studies were made to understand the transition in triangulated surface models \cite{KANTOR-NELSON-PRA1987,WHEATER-NPB1996,BCFTA-JP96-NPB9697,Baum-Ho-PRA1990,CATTERALL-PLB1989,CATTERALL-NPBSUP1991,AMBJORN-NPB1993,ABGFHHM-PLB1993,BCHHM-NPB9393,KOIB-PLA200234,KOIB-EPJB-20056,KOIB-PRE-20034}. The transition was reported as first-order in recent numerical studies \cite{KD-PRE2002,KOIB-PRE-20045-NPB-2006}. 

On the other hand, the concern with inhomogeneous surfaces has been growing over the past decade \cite{MSWD-PRE-1994,JSWW-PRE-1995}. A homogeneous artificial membrane that is coated by elastic skeletons is also considered to be an inhomogeneous membrane. Some of the mechanical properties of such membranes were revealed experimentally \cite{HHBRM-PRL-2001}. The hop diffusion of membrane protein or lipids was observed, and as a consequence the compartment of cytoskeletons was confirmed to be in biological membranes \cite{Kusumi-BioJ-2004}. It is also well known that the microtubule, which is an element of the cytoskeleton, gives a mechanical strength to the surface of the biological membranes.  

However, the surface collapsing phenomena and the surface fluctuation phenomena are almost unknown in such inhomogeneous models for membranes. Therefore, it is worthwhile to study an inhomogeneous fluid surface model within the framework of the conventional surface model of HPK. We note that the inhomogeneity in our model corresponds to the cytoskeletons in biological membranes as stated above. The fluidity realized by dynamical triangulations in the inhomogeneous model, as well as the fluidity in the homogeneous surface models, corresponds to the lateral diffusion of lipids in membranes. 

In this paper we study a compartmentalized surface model by Monte Carlo (MC) simulations. The Hamiltonian of the model includes no two-dimensional bending energy but a one-dimensional bending energy. The model is defined on dynamically triangulated surfaces, where the free diffusion of vertices is confined inside the compartments. The mechanical strength of the surface is given only by the compartment boundary, which is composed of one-dimensional elastic chains and rigid junctions. Because the collapsed phase is expected to appear at sufficiently small bending rigidity $b[kT]\!\to\! 0 (b\!\not=\!0)$, we concentrate on the phase structure at relatively large $b$ in this paper. Consequently, information on the phase boundary at $b\!\to\! 0$ remains unanswered.

We recently reported numerical results of three types of surface models \cite{KOIB-2007,KOIB-JSTP2007}, which are similar to the model in this paper. Then, we should comment on the similarity/difference between the model in this paper and the models in \cite{KOIB-2007,KOIB-JSTP2007}. Firstly, the lattice structure of the model in this paper is very similar to that of the first model in \cite{KOIB-2007} and that of the model in \cite{KOIB-JSTP2007}, and is identical to that of the second model in \cite{KOIB-2007}. Secondly, the lattice in this paper and that of the first model in \cite{KOIB-2007} are the dynamically triangulated one, while the lattice of the second model in \cite{KOIB-2007} and that in \cite{KOIB-JSTP2007} are the fixed-connectivity one. Thirdly, the Hamiltonian is different from the one in the first model in \cite{KOIB-2007}. The Hamiltonian of the model of this paper includes only one-dimensional bending energy, which is defined on the compartment boundary, while the Hamiltonian of the first in \cite{KOIB-2007} includes only a two-dimensional bending energy, which is defined all over the surface, and no one-dimensional bending energy is given to the compartment boundary. Therefore, the model in this paper is different from the three models in \cite{KOIB-2007,KOIB-JSTP2007}. 

Our results obtained in this paper show that the model undergoes a first-order transition between the smooth phase and the crumpled phase. Moreover, the smooth phase can be divided into the spherical phase and the planar phase, and the crumpled phase can also be divided into the tubular phase and the collapsed phase, which is expected to appear at sufficiently small $b$ because no self-avoiding property  \cite{GREST-JPIF1991,BOWICK-TRAVESSET-EPJE2001,BCTT-PRL2001} is assumed in the model. It must be emphasized that such variety of phases can be seen neither in the conventional surface models nor in the compartmentalized models such as those in \cite{KOIB-2007,KOIB-JSTP2007}.

One remarkable result is the appearance of planar surfaces. The echinocytic shapes of erythrocytes were extensively studied, and they are currently known to be described by many models such as the area difference bilayer model \cite{JSWW-PRE-1995}. The shape of membranes is also sensitive to the flow fields \cite{NOGUCHI-GOMPPER-PRL2004}. Our model in this paper indicates that one possible origin of such planar shape comes from the inhomogeneity due to the cytoskeltal structure and the fluidity of lateral diffusion of vertices.      

\section{Model}\label{model}
\begin{figure}[htb]
\unitlength 0.1in
\begin{picture}( 0,0)(  10,10)
\put(15,8.5){\makebox(0,0){(a) $(N,N_S,N_J,L)\!=$ }}%
\put(15.7,7){\makebox(0,0){$(2322,600,42,6)$ }}%
\put(33,8.6){\makebox(0,0){(b) A rigid junction}}%
\put(33.7,7.1){\makebox(0,0){with the chains }}%
\end{picture}%
\vspace{0.5cm}
\centering
\includegraphics[width=3.5cm]{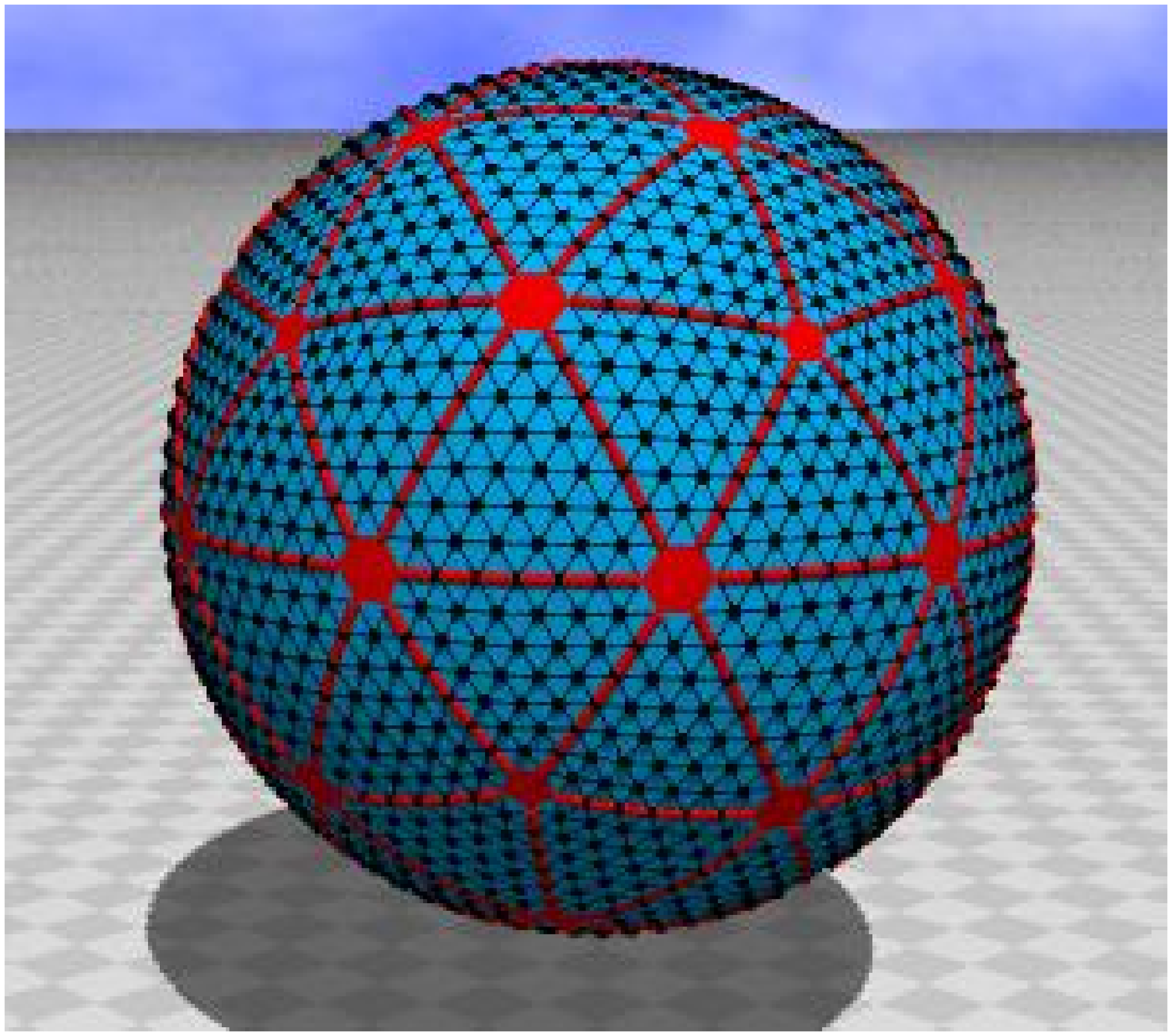} 
\includegraphics[width=4.5cm]{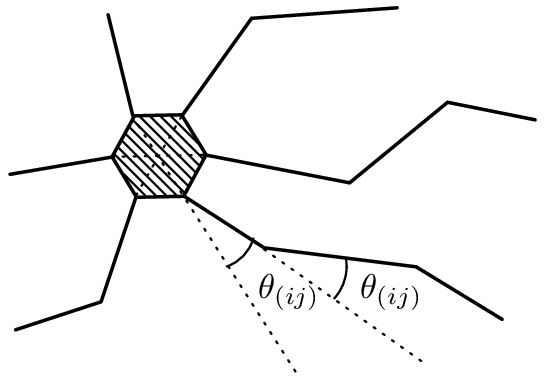}
\caption{(Color online) (a) Starting configuration of surfaces of size $(N,N_S,N_J,L)\!=\!(2322,600,42,6)$, and (b) angles $\theta_{(ij)}$ in the  bending energy $S_2$ of Eq.(\ref{Disc-Eneg}). Thick lines in (a) denote the compartment boundary composed of the linear chains and the rigid junctions of the hexagonal and the pentagonal plates, whose size is drawn many times larger than that of the lattices for the simulations.}
\label{fig-1}
\end{figure}
Figure \ref{fig-1}(a) shows a triangulated surface of size $(N,N_S,N_J,L)\!=\!(2322,600,42,6)$, where $N$ is the total number of vertices including the junctions, $N_S$ is the total number of vertices on the chains, $N_J$ is the total number of junctions, and $L$ is the length of chains between the two nearest-neighbor junctions. It should be noted again that $N_J$ is included in $N$; junctions are counted in the total number of vertices. The junctions are assumed as rigid plates; twelve of them are pentagon and all the others are hexagon. The junction size in Fig.\ref{fig-1}(a) is drawn many times larger than that of the lattices for the simulations; and it will be discussed in the last part of this section. Thick lines on the surface in Fig.\ref{fig-1}(a) denote the chains, which are terminated at the junctions.

 The construction of the lattices is as follows: Let us start with the icosahedron. Every edge of the icosahedron is divided into $\ell$ pieces of uniform length, and then we have a triangulated surface of size $N_0\!=\!10\ell^2\!+\!2$ (= the total number of vertices on the surface). The compartmentalized structures are constructed by dividing $\ell$ further into $m$ pieces ($m\!=\!1,2,\cdots$). Thus, we have the chains of uniform length $L\!=\!(\ell /m)\!-\!2$ when $m$ divides $\ell$. The reason for the subtraction $-2$ is because of the junctions at the two end points of the chain. Because the compartmentalized structure is a sublattice, the total number of junctions $N_J$ is given by $N_J\!=\!10m^2\!+\!2$. The total number of bonds in the sublattice is $3N_J\!-\!6$, and each bond contains $L\!-\!1$ vertices, then $N_S$ is given by $N_S\!=\!(3N_J\!-\!6)(L\!-\!1)$, which can be written as $N_S\!=\!30m(\ell\!-\!3m)$. The hexagonal (pentagonal) rigid junctions are composed of $7$ ($6$) vertices, then $N_J\!-\!12$ hexagonal rigid junctions and   $12$ pentagonal rigid junctions reduce the total number of vertices $N_0$ by $(N_J\!-\!12)\!\times\! 6$ and $12\!\times\! 5$. Therefore, we have $N\!=\!N_0\!-\!6N_J\!+\!12$, which can also be written as $N\!=\!10\ell^2\!-\!60m^2\!+\!2$. The thermodynamic limit of our model is defined by $N\!\to\!\infty$, $N_S\!\to\!\infty$, and $N_J\!\to\!\infty$  under the condition that $L$ is finite. We have the thermodynamic limit at $\ell\!\to\! \infty$ and $m\!\to\!\infty$. The lattice of size $(N,N_S,N_J,L)\!=\!(2322,600,42,6)$ in Fig.\ref{fig-1}(a) is given by two independent integers $(\ell,m)\!=\!(16,2)$.

The surfaces can be characterized by the length $L$. In this paper, we assume three values for $L$ such that
\begin{equation}
\label{Length} 
L=6,\quad L=8,\quad L=11.
\end{equation} 
The value of $L$ has a one to one correspondence with the total number of vertices $n$ in a compartment; in fact, the values of $L$ in Eq.(\ref{Length}) correspond to $n\!=\!21$, $n\!=\!36$, and $n\!=\!66$, respectively \cite{KOIB-2007}. We note that the effective physical meaning of increasing (decreasing) $L$ can be considered as the increasing (decreasing) temperature. In fact, the surface fluctuation mainly comes from the thermal fluctuation of vertices inside the compartments. Because no bending energy is assumed inside the compartments, the fluctuation of vertices becomes large not only in the in-plane directions (free diffusion) but also in the direction perpendicular to the surface. Thus, we consider that the fluctuations are expected to grow with increasing $n$, i.e., increasing $L$.

 We use the surfaces of size $(N,N_S,N_J)$ listed in Table \ref{table-1}. Three different sizes $(N,N_S,N_J)$ are assumed for each $L$. The corresponding integers $(\ell,m)$ are as follows: $(16,2)$,  $(24,3)$, and  $(32,4)$ for the $L\!=\!6$ surfaces,  $(10,1)$,  $(20,2)$, and  $(30,3)$ for the $L\!=\!8$ surfaces, and $(13,1)$,  $(26,2)$, and  $(39,3)$ for the $L\!=\!11$ surfaces.
\begin{table}[hbt]
\caption{The surface size assumed in the simulations. Three sizes $(N,N_S,N_J)$ are assumed for each $L$.}
\label{table-1}
\begin{center}
 \begin{tabular}{cccc}
$L$  & $(N,N_S,N_J)$ & $(N,N_S,N_J)$  & $(N,N_S,N_J)$  \\
 \hline
  6  & (2322,600,42)  & (5222,1350,92) & (9282,2400,162)  \\
  8  & (942,210,12) & (3762,840,42) & (8462,1890,92)   \\
  11  & (1632,300,12) & (6522,1200,42) & (14672,2700,92)   \\
 \hline
 \end{tabular} 
\end{center}
\end{table}

The model is defined by the partition function 
\begin{eqnarray} 
\label{Part-Func}
 Z = \sum_{\cal T} \int^\prime \prod _{i=1}^{N} d X_i \exp\left[-S(X,{\cal T})\right],\\  
 S(X,{\cal T})=S_1 + b S_2, \nonumber
\end{eqnarray} 
where $S_1$ is the Gaussian bond potential, which is defined all over the surface, and $S_2$ is the one-dimensional bending energy, which is defined on the compartment boundary and will be given below. The parameter $b$ is the bending rigidity. The integration symbol $\int^\prime$ in Eq.(\ref{Part-Func}) denotes that the center of mass of the surface is fixed. $\sum_{\cal T}$ denotes the sum over all possible triangulations ${\cal T}$, which are performed by the bond flip technique keeping the compartments unflipped. The bond flip procedure will be given in the following section. 

 The integration measure $\prod _{i=1}^{N} d X_i$ is given by the product
\begin{eqnarray}
\label{Measure}
&&\prod _{i=1}^{N} d X_i=\left(\prod _{i=1}^{N'} d X_i q_i^{\alpha}\right)\left( \prod _{i=1}^{N_J} d X_i \prod _{j(i)} q_{j(i)}^{\alpha}\right), \nonumber \\
&&(\alpha=3/2, \;  \;0),
\end{eqnarray} 
 where $N^\prime$ ($\!=\!N\!-\!N_J$) is the total number of vertices excluding the junctions,  $\prod _{i=1}^{N'} d X_i q_i^{\alpha} $ denotes the integration over the $3D$ translational degrees of freedom (DOF) of the vertices $i$, and $\prod _{i=1}^{N_J} d X_i\prod _{j(i)} q_{j(i)}^{\alpha}$ denotes those of the $3D$ translational DOF and the $3D$ rotational DOF of the junctions $i$. The co-ordination number $q_i$ is the total number of bonds meeting at the vertex $i$,  and $q_{j(i)}$ is the total number of bonds meeting at the corner $j(i)$ of the junction $i$. 

The parameter $\alpha$ was chosen to be $\alpha\!=\!3/2$ in \cite{David-NPB1985,BKKM-NPB1986}, while $\alpha\!=\!0$ in many previous simulations on dynamically triangulated surfaces in the literatures. It is easy to understand that large positive $\alpha$ suppresses the configurations with large coordination number. Therefore, it is interesting to see the dependence of the phase structure on $\alpha$. 
  
 We chose both $\alpha\!=\!3/2$ and $\alpha\!=\!0$ for the weight $q_i^\alpha$ \cite{David-NPB1985,BKKM-NPB1986}, and see whether the phase structure of the model depends on $\alpha$ or not. If the parameter is chosen to $\alpha\!=\!3/2$, then the coordination number $q_i$ serves as a weight of the integration $d X_i$, while $\alpha\!=\!0$ gives the uniform weight. The weight $\prod _{i=1}^{N'} q_i^{\alpha} $ can also be written as $\prod _{i=1}^{N'} q_i^{\alpha} \!=\!\exp(\alpha \sum_i \log q_i)$, and therefore,  $\prod _{i=1}^{N'} q_i^{\alpha} $ is considered to be the co-ordination dependent term $-\alpha \sum_i \log q_i$ in the Hamiltonian; $-\alpha \sum_i \log q_i$ changes its value only on dynamically triangulated surfaces.

The Gaussian term $S_1$ and the bending energy term $S_2$ are defined by
\begin{equation}
\label{Disc-Eneg} 
S_1=\sum_{(ij)} \left(X_i-X_j\right)^2,\quad S_2=\sum_{(ij)} \left[ 1-\cos \theta_{(ij)} \right],
\end{equation} 
where $\sum_{(ij)}$ in $S_1$ is the sum over bonds $(ij)$ connecting the vertices $i$ and $j$, and $\sum_{(ij)}$ in $S_2$ is also the sum over bonds $(ij)$. $\theta_{(ij)}$ in  $S_2$ is the angle between the bonds $i$ and $j$, which include {\it virtual} bonds. The {\it virtual} bonds denote the lines between the center and the corners of the junction; the hexagonal (pentagonal) junction contains six (five) virtual bonds. 

 Figure \ref{fig-1}(b) is a junction and the chains linked to the junction on a fluctuating surface. Triangles are eliminated from the figure. One $\theta_{(ij)}$ shown at a corner of the junction is defined by using a virtual bond and a real bond in a chain, and the other  $\theta_{(ij)}$ shown at a vertex is defined by real bonds on the same chain. 

The size of the junctions can be characterized by the edge length $R$, which is fixed to 
\begin{equation}
\label{Edge-Length} 
R=0.1\quad({\rm edge\; length\; of\; the \; junctions} ).
\end{equation}
The value $R\!=\!0.1$ is relatively smaller than the mean bond length $0.707$, which corresponds to the relation $S_1/N\!=\!1.5$ satisfied in the equilibrium configuration of surfaces without the rigid junctions. As we will see later, the relation $S_1/N\!=\!1.5$ is slightly violated in the model of this paper because of the rigid junctions. 

Here we comment on the unit of physical quantities. Let $a$ be the length scale of the model, then the unit of physical quantity that has the length unit can be expressed by $a$; the unit of $S_1$ is $[a^2]$. The surface tension coefficient $\lambda$ in $\lambda S_1 \!+\! bS_2$ has the unit $[kT/a^2]$ and assumed to be $\lambda\!=\!1 [kT/a^2]$, and the bending rigidity $b$ has the unit of $[kT]$  as described above. 

Note that the bending rigidity $b$ in the Hamiltonian is a microscopic quantity from the view point of statistical mechanical model, and therefore $b$ is not always identical to the macroscopic bending rigidity of real physical membranes. However, the microscopic value $b$ of real membranes can effectively be varied with the temperature, because $b$ has the unit of $kT$. Therefore, it is possible to consider that the phase structure described in terms of $b$ in the surface model corresponds to the phase structure described in terms of $T$ in real physical membranes. The length scale $a$ in the model is also a microscopic quantity and, we consider that $a$ is sufficiently smaller than the membrane size. 

\section{Monte Carlo technique}\label{MC-Techniques}
A sequence of random numbers called Mersenne Twister \cite{Matsumoto-Nishimura-1998} is used in the canonical MC simulations. The Metropolis technique is applied to update $X$ and ${\cal T}$, where the variable $X$ denotes the position of the vertices and that of the junctions. The vertex position $X$ is shifted so that $X^\prime \!=\! X\!+\!\delta X$, where $\delta X$ is randomly chosen in a small sphere. The new position $X^\prime$ is accepted with the probability ${\rm Min}[1,\exp(-\Delta S)]$, where $\Delta S\!=\! S({\rm new})\!-\!S({\rm old})$. The position $X$ of a hexagonal (or pentagonal) junction, which is not a point but a rigid plate, is also integrated out by performing $3D$ random translations and $3D$ random rotations.

Thus, the variable $X$ is updated by a random $N^\prime$ ($\!=\!N\!-\!N_J$) shifts of vertices, a random $N_J$ translations of junctions, and  a random $N_J$ rotations of junctions. These updates are denoted by $(N^\prime,N_J,N_J)$ updates of $X$. The $N^\prime$ shifts of $X$ can be divided into $N_S$ shift of the vertices on the linear chains and $N^\prime\!-\!N_S$ shifts of all the other vertices, which are those inside the compartments. 

The radius of the small sphere for $\delta X$ is fixed at the beginning of the MC simulations in order to maintain about $50\%$ acceptance rate. The vertices on the linear chains carry the bending energy $S_2$ in Eq.(\ref{Disc-Eneg}), while  all the other vertices inside the compartments does not. Therefore, the acceptance rate is independently controlled in the two-groups of vertices. The radius for the random translation of the junctions and that for the random rotation are also independently chosen so that the acceptance rates are both about $50\%$. 

The summation over ${\cal T}$ in $Z$ of Eq.(\ref{Part-Func}) is performed by using the standard bond flip technique \cite{Baum-Ho-PRA1990,CATTERALL-PLB1989}. The flip is accepted with the probability ${\rm Min} [1, \exp(-\Delta S)]$. The acceptance rate for the bond flip is not under control and is about $75\%$, which is almost independent of $b$.

The bonds are labeled with sequential numbers. The total number of bonds is denoted by $N_B^\prime$, which excludes the number of bonds on the linear chains because the bonds on the linear chains remain unflipped. 

The bond flip is performed as follows: Firstly, the odd-numbered bonds are sequentially chosen to be flipped for the $N_B^\prime/2$ updates of ${\cal T}$, and after that, the $(N,N_J,N_J)$ updates of $X$ are performed. Secondly, the remaining even-numbered bonds are chosen to be flipped for the $N_B^\prime/2$ updates of ${\cal T}$, and after that, the $(N,N_J,N_J)$ updates of $X$ are performed. Thus, the $(N,N_J,N_J)$ updates of $X$ and the $N_B^\prime/2$ updates of ${\cal T}$ are consecutively performed, and these make one MCS (Monte Carlo Sweep).

 We introduce the lower bound $1\times 10^{-8}$ to the area of triangles. No lower bound is imposed on the bond length.  

\section{Results of simulation}\label{Results}
\subsection{$\alpha=3/2$}\label{alpha3/2}
As mentioned in Section \ref{model}, we assume the value of $\alpha$ in Eq.(\ref{Measure}) as $\alpha\!=\!3/2$ and $\alpha\!=\!0$. In this subsection, we present the results obtained under $\alpha\!=\!3/2$ by using snapshots and figures, and in the next subsection we will show some of the results under $\alpha\!=\!0$.  

The thermalization MCS is $1\times 10^7$ in almost all cases. However, more than $1\times 10^8$ thermalization MCS were done close to the transition point in such cases that the surface is trapped in one phase at first and then changes its phase to a more stable one under a given condition. The total number of MCS for the production of samples is $0.8\times 10^8\sim1.3\times 10^8 $. At the transition point, about $2\times 10^8$ MCS was performed after the thermalization in some cases.    

\begin{figure}[htb]
\unitlength 0.1in
\begin{picture}( 0,0)(  10,10)
\put(14,37.5){\makebox(0,0){(a) $b\!=21.2$ }}%
\put(25,37.5){\makebox(0,0){(b) $b\!=21.8$ }}%
\put(35.5,37.5){\makebox(0,0){(c) $b\!=22$ }}%
\put(15,9){\makebox(0,0){(d) The section }}%
\put(26,9){\makebox(0,0){(e) The section  }}%
\put(37,9){\makebox(0,0){(f) The section  }}%
\end{picture}%
\vspace{0.3cm}
\includegraphics[width=8.5cm]{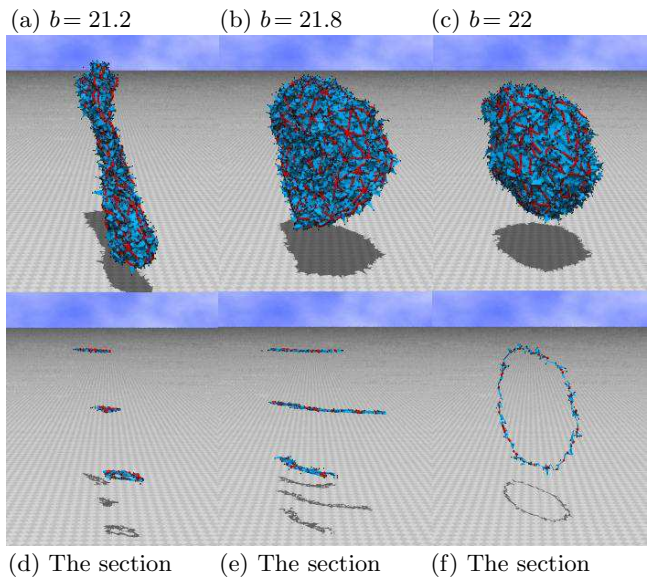}
\caption{(Color online) The snapshots of surfaces of size $(N,N_S,N_J,L)\!=\!(8462,1890,92,8)$ obtained at (a) $b\!=\!21.2$ (tubular phase), (b) $b\!=\!21.8$ (planar phase), and (c) $b\!=\!22$ (spherical phase), and  (d),(e),(f) are the surface sections of (a),(b),(c), respectively. $\alpha\!=\!3/2$. } 
\label{fig-2}
\end{figure}
We show snapshots of the $(N,N_S,N_J,L)\!=\!(8462,1890,92,8)$ surface in Figs.\ref{fig-2}(a)--\ref{fig-2}(c). They were obtained at (a) $b\!=\!21.2$, (b) $b\!=\!21.8$, and (c) $b\!=\!22$, which respectively corresponds to the tubular phase, the planar phase, and the spherical phase. 
The snapshot of Fig.\ref{fig-2}(b) at $b\!=\!21.8$ was the final configuration produced after $2\times 10^8$ MCS including $1\times 10^8$ thermalizaion MCS; the planar surface was stable after the thermalization MCS. The surface sections are shown in Figs.\ref{fig-2}(d)--\ref{fig-2}(f); the sections in Figs.\ref{fig-2}(d) and \ref{fig-2}(e) were obtained by slicing the surfaces perpendicular to the vertical axis, and the section in Figs.\ref{fig-2}(f) was obtained by slicing the surface perpendicular to a horizontal axis. All of the snapshots were drawn in the same scale. The axis of the tubular surface Fig.\ref{fig-2}(a) as well as the axis perpendicular to the planar surface Fig.\ref{fig-2}(b) is spontaneously chosen.     

The planar phase is stable only on the $L\!=\!8$ surfaces, while it seems unstable on the $L\!=\!6$ surfaces and on the $L\!=\!11$ surfaces. Even if the planar phase once appears on the surfaces of $L\!=\!6$ and $L\!=\!11$ of size at least $N\!\leq\! 9282$ and $N\!\leq\! 14672$, respectively, it eventually collapses into the tubular phase. Therefore, we find that no planar phase can be seen on the $L\!=\!6$ and the $L\!=\!11$ surfaces; the tubular phase and the spherical phase are connected by a discontinuous transition on those surfaces. Thus, we understand that the planar phase appears depending on the size of the compartments. We should note that the planar surface may bend and fluctuate in the limit of $N\!\to\!\infty$, and the tubular surface may also bend and wind in the same limit.

\begin{figure}[htb]
\centering
\includegraphics[width=8.5cm]{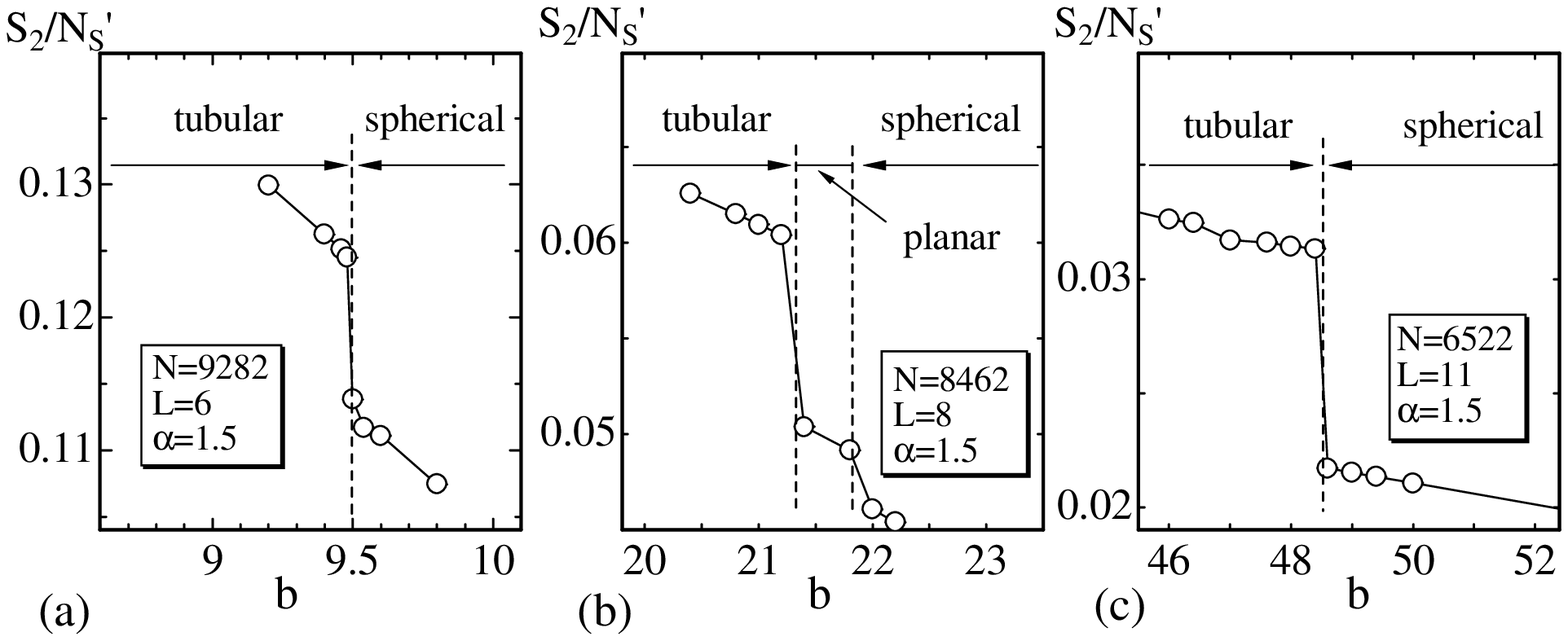}
\caption{The one-dimensional bending energy $S_2/N_S^\prime$ against $b$ obtained on the surfaces of (a) $L\!=\!6$, (b) $L\!=\!8$, and (c) $L\!=\!11$. $N_S^\prime(\!=\!N_S\!+\!6N_J\!-\!12)$ is the total number of vertices where $S_2$ is defined.  } 
\label{fig-3}
\end{figure}
Figures \ref{fig-3}(a),\ref{fig-3}(b), and \ref{fig-3}(c) show the bending energy $S_2/N_S^\prime$ of Eq.(\ref{Disc-Eneg}) against $b$, which were obtained on the surfaces of $L\!=\!6$, $L\!=\!8$, and $L\!=\!11$, respectively. $N_S^\prime(\!=\!N_S\!+\!6N_J\!-\!12)$ is the total number of vertices where $S_2$ is defined. $6N_J\!-\!12$ is the total number of corners of the junctions, which include $12$-pentagons. The solid lines on the data were drawn to guide the eyes. Dashed lines drawn vertically denote the phase boundary between the tubular and the spherical phases, the boundary between the tubular and the planar phases, and the boundary between the planar and the spherical phases. The discontinuous change of $S_2/N_S^\prime$ between the tubular phase and the spherical (or the planar) phase is very clear in the figures and considered to be a sign of the first-order transition.

\begin{figure}[htb]
\centering
\includegraphics[width=8.5cm]{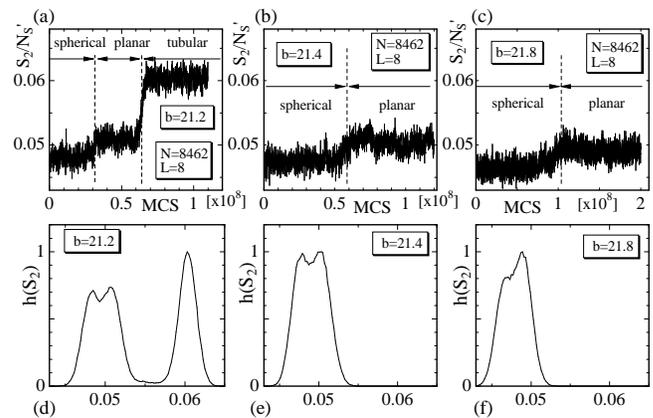}
\caption{The variation of $S_2/N_S^\prime$ against MCS, which were obtained on the $(N,N_S,N_J,L)\!=\!(8462,1890,92,8)$ surface at (a) $b\!=\!21.2$, (b) $b\!=\!21.4$, and (c) $b\!=\!21.8$. The dashed lines denote the MCS where the jumps occurred. The corresponding normalized histogram $h(S_2)$ obtained at (d) $b\!=\!21.2$, (e) $b\!=\!21.4$, and (f) $b\!=\!21.8$. The parameter $\alpha$ was fixed to $\alpha\!=\!3/2$. } 
\label{fig-4}
\end{figure}
In order to see the difference between $S_2/N_S^\prime$ in those three phases, we plot in Figs.\ref{fig-4}(a),\ref{fig-4}(b), and \ref{fig-4}(c) the variation of $S_2/N_S^\prime$ against MCS obtained at $b\!=\!21.2$,  $b\!=\!21.4$, and $b\!=\!21.8$ on the $(N,N_S,N_J,L)\!=\!(8462,1890,92,8)$ surface. The thermalization MCS were not discarded; they were included only in those variations. $S_2/N_S^\prime$ at $b\!=\!21.2$ in Fig.\ref{fig-4}(a) shows a jump from the spherical phase to the planar phase and a jump from the planar phase to the tubular phase; the corresponding MCS at the jumps were indicated with the dashed vertical lines. We also find in Fig.\ref{fig-4}(b) a jump from the spherical phase to the planar phase. A jump is also seen in $S_2/N_S^\prime$ at $b\!=\!21.8$ in Fig.\ref{fig-4}(c) from the spherical phase to the planar phase. 

The value of $b\!=\!21.2$ corresponds to the tubular phase, whereas $b\!=\!21.4$ and $b\!=\!21.8$ correspond to the planar phase, because the final states are considered to be stable states. The surfaces at $b\!=\!21.2$ and $b\!=\!21.8$ can be seen in the snapshots in Figs.\ref{fig-3}(a) and \ref{fig-3}(b). 

The distribution of $S_2/N_S^\prime$ are shown as the normalized histograms $h(S_2)$ in Figs.\ref{fig-4}(d)--\ref{fig-4}(f), which respectively correspond to the variations in Figs.\ref{fig-4}(a)--\ref{fig-4}(c). We see that $h(S_2)$ in Fig.\ref{fig-4}(d) has three peaks; two of them are almost overlapping and the other one is distinctly separated from the previous two. Those three peaks in $h(S_2)$  correspond to the spherical phase, planar phase, and the tubular phase. Two almost overlapping peaks can also be seen in $h(S_2)$ in Figs.\ref{fig-4}(e) and \ref{fig-4}(f), and they are corresponding to the spherical phase and the planar phase. We remark that the surfaces hardly change not only from the tubular phase to the smooth (= spherical or planar) phase but also from the planar phase to the spherical phase on the $L\!=\!8$ and $L\!=\!11$ surfaces. For this reason, we find in Figs.\ref{fig-4}(a)--\ref{fig-4}(c) no jump-back from a higher $S_2$ state (such as the tubular state) to a lower $S_2$ state (such as the planar state). 

\begin{figure}[htb]
\centering
\includegraphics[width=8.5cm]{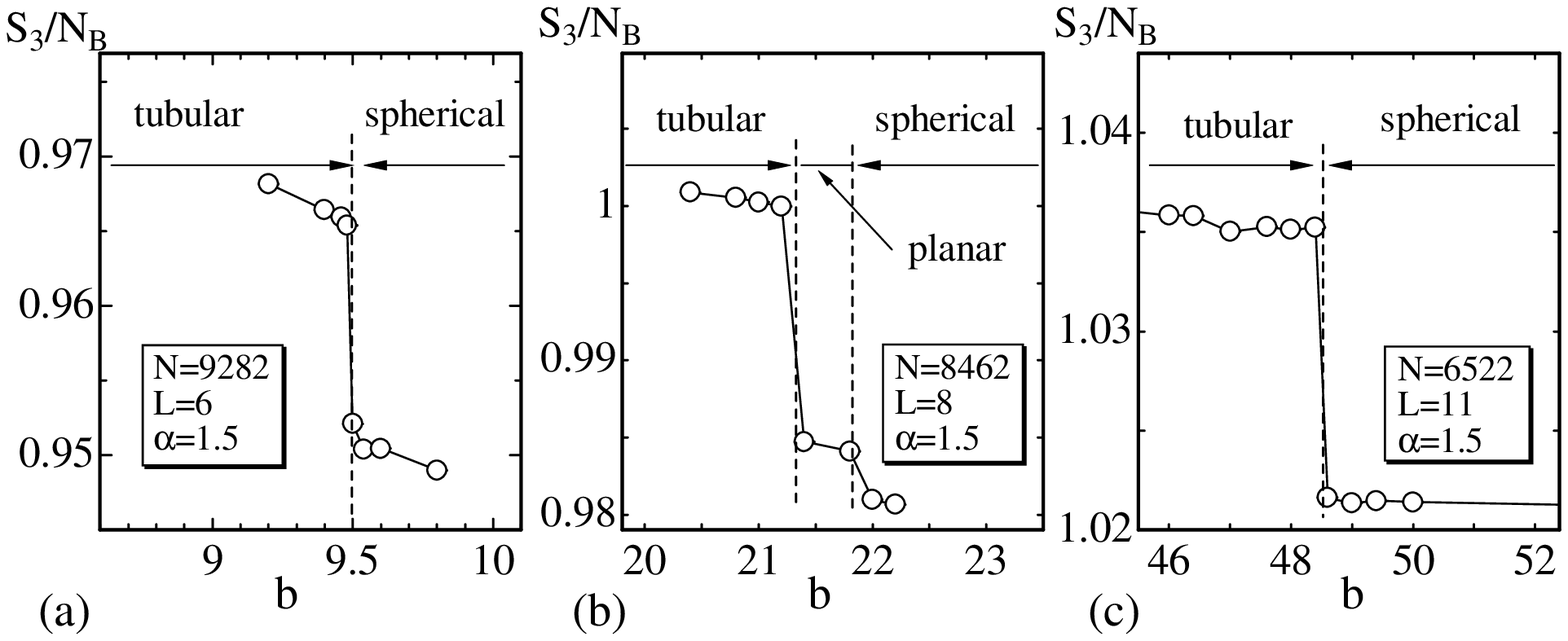}
\caption{The two-dimensional bending energy $S_3/N_B$ against $b$ obtained on the surfaces of (a) $L\!=\!6$, (b) $L\!=\!8$, and (c) $L\!=\!11$. $N_B$ is the total number of bonds where $S_3$ is defined. } 
\label{fig-5}
\end{figure}
The two-dimensional bending energy is defined by
\begin{equation}
S_3=\sum_{(ij)}\left(1-{\bf n}_i\cdot {\bf n}_j\right), 
\end{equation}
where ${\bf n}_i$ is the unit normal vector of the triangle $i$, and ${\bf n}_i\cdot {\bf n}_j$ is defined on the common bond $(ij)$ of the triangles $i$ and $j$. $S_3$ is not included in the Hamiltonian and is defined even on the edges of the rigid junctions. Figures \ref{fig-5}(a)--\ref{fig-5}(c) show $S_3/N_B$ against $b$ obtained on the surfaces of $L\!=\!6$, $L\!=\!8$, and $L\!=\!11$, where $N_B$ is the total number of bonds including the edges of the junctions. The jump of $S_3/N_B$ in Fig.\ref{fig-5}(b) is clearly seen between the tubular phase and the planar phase. On the contrary, $S_3/N_B$ in the planar phase in Fig.\ref{fig-5}(b), as well as $S_2/N_S^\prime$ in the planar phase in Fig.\ref{fig-3}(b), is not so clearly distinguishable from that in the spherical phase.   

\begin{figure}[htb]
\centering
\includegraphics[width=8.5cm]{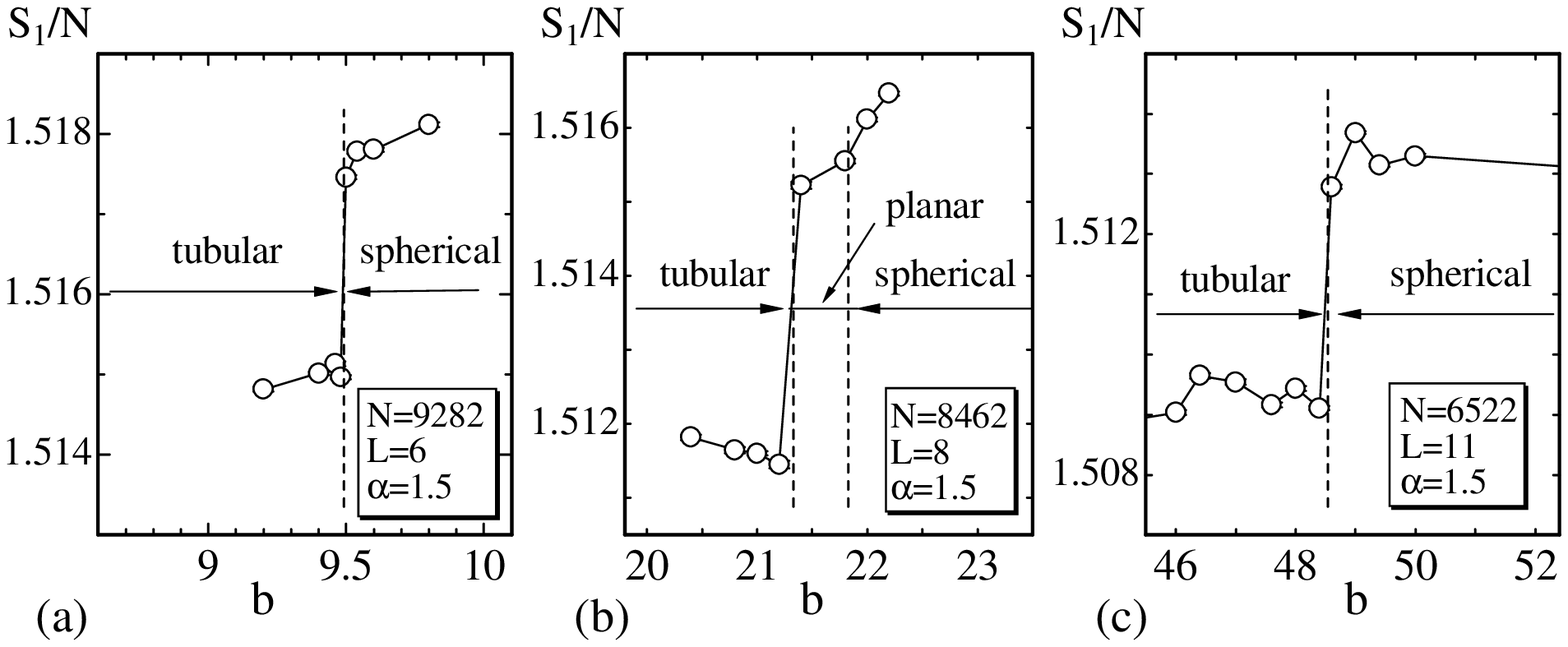}
\caption{The Gaussian bond potential $S_1/N$ against $b$ obtained on the surfaces of (a) $L\!=\!6$, (b) $L\!=\!8$, and (c) $L\!=\!11$. } 
\label{fig-6}
\end{figure}
It is expected that the Gaussian bond potential $S_1/N$ is influenced by the phase transitions. The potential $S_1/N$ should be $S_1/N\simeq 3/2$, which is satisfied in the model without the rigid junctions because of the scale invariant property of the partition function in that case. However, the junction size $R$ in Eq.(\ref{Edge-Length}) is finite in the model of this paper, and therefore $S_1/N$ can slightly deviate from $3/2$.

Figures \ref{fig-6}(a)--\ref{fig-6}(c) show $S_1/N$ against $b$ obtained on the surfaces of (a) $L\!=\!6$, (b) $L\!=\!8$, and (c) $L\!=\!11$. Discontinuous changes in $S_1/N$ shown in the figures are consistent with the discontinuous transitions of the model, although the changes are very small compared to the value of $S_1/N$ itself. We find also the expected deviation of $S_1/N$ from $3/2$ in the figures.

\begin{figure}[htb]
\centering
\includegraphics[width=8.5cm]{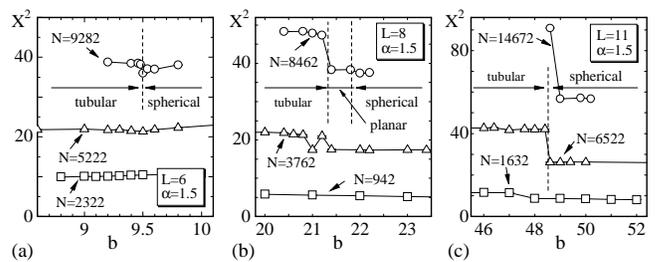}
\caption{The mean square size $X^2$ against $b$ obtained on the surfaces of (a) $L\!=\!6$, (b) $L\!=\!8$, and (c) $L\!=\!11$. } 
\label{fig-7}
\end{figure}
Figures \ref{fig-7}(a)--\ref{fig-7}(c) show the mean square size $X^2$, which is defined by
\begin{equation}
\label{Mean-Square-Size}
X^2= {1\over N} \sum _i \left( X_i-\bar X \right)^2,\quad \bar X = {1\over N} \sum_i X_i,
\end{equation}
where $\bar X$ is the center of mass of the surface. We see that the phase transition is not reflected in $X^2$ on the $L\!=\!6$ surfaces in Fig.\ref{fig-7}(a), and the transition is also not reflected in $X^2$ on the $L\!=\!8$ surfaces in Fig.\ref{fig-7}(b) at the transition point between the planar phase and the spherical phase. To the contrary, $X^2$ discontinuously changes in Fig.\ref{fig-7}(b) at the transition point between the tubular phase and the planar phase and also at the transition point in Fig.\ref{fig-7}(c). All of these behaviors of $X^2$ at the transition points are consistent with those of $S_2/N_S^\prime$, $S_3/N_B$, and $S_1/N$.  

\subsection{$\alpha=0$}\label{alpha0}
In this subsection, we present some of the results obtained under $\alpha\!=\!0$.

\begin{figure}[htb]
\unitlength 0.1in
\begin{picture}( 0,0)(  10,10)
\put(14,34.5){\makebox(0,0){(a) $b\!=20.9$ }}%
\put(25,34.5){\makebox(0,0){(b) $b\!=21.4$ }}%
\put(35.5,34.5){\makebox(0,0){(c) $b\!=21.8$ }}%
\put(15,9){\makebox(0,0){(d) The section }}%
\put(26,9){\makebox(0,0){(e) The section  }}%
\put(37,9){\makebox(0,0){(f) The section  }}%
\end{picture}%
\vspace{0.3cm}
\includegraphics[width=8.5cm]{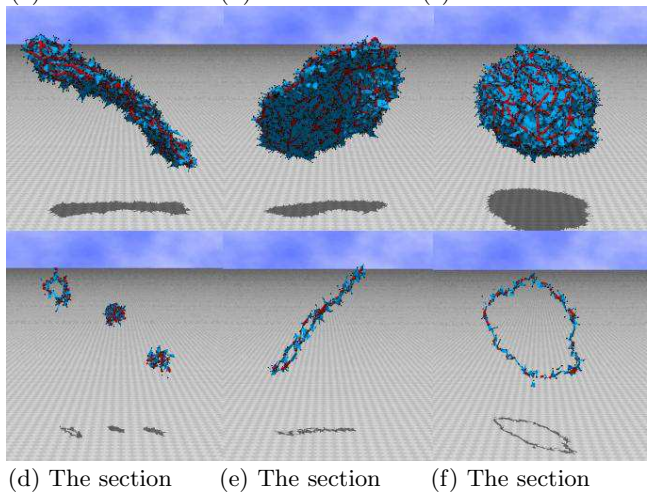}
\caption{(Color online) The snapshots of surfaces of size $(N,N_S,N_J,L)\!=\!(8462,1890,92,8)$ obtained at (a) $b\!=\!20.9$ (tubular phase), (b) $b\!=\!21.4$ (planar phase), and (c) $b\!=\!21.8$ (spherical phase), and  (d),(e),(f) are the surface sections of (a),(b),(c), respectively. $\alpha\!=\!0$. } 
\label{fig-8}
\end{figure}
Snapshots of surfaces of $\alpha\!=\!0$ are shown in Figs.\ref{fig-8}(a), \ref{fig-8}(b), \ref{fig-8}(c), which respectively correspond to the tubular phase ($b\!=\!20.9$),  the planar phase ($b\!=\!21.4$), and the spherical phase ($b\!=\!21.8$). The surface size is $(N,N_S,N_J,L)\!=\!(8462,1890,92,8)$, which is identical to that in Fig.\ref{fig-2}. The snapshot in Fig.\ref{fig-8}(b) at $b\!=\!21.4$ is the final configuration produced after $1.9\times 10^8$ MCS including $1\times 10^7$ thermalizaion MCS; the planar surface was stable throughout the simulation. Thus, we find that three distinct phases are seen also in the surfaces of $L\!=\!8$, and that the planar phase is unstable on the surfaces of $L\!=\!6$ and $L\!=\!11$ under the condition $\alpha\!=\!0$. Therefore, we consider that the phase structure of the model is independent of whether $\alpha\!=\!3/2$ or $\alpha\!=\!0$.

\begin{figure}[htb]
\centering
\includegraphics[width=8.5cm]{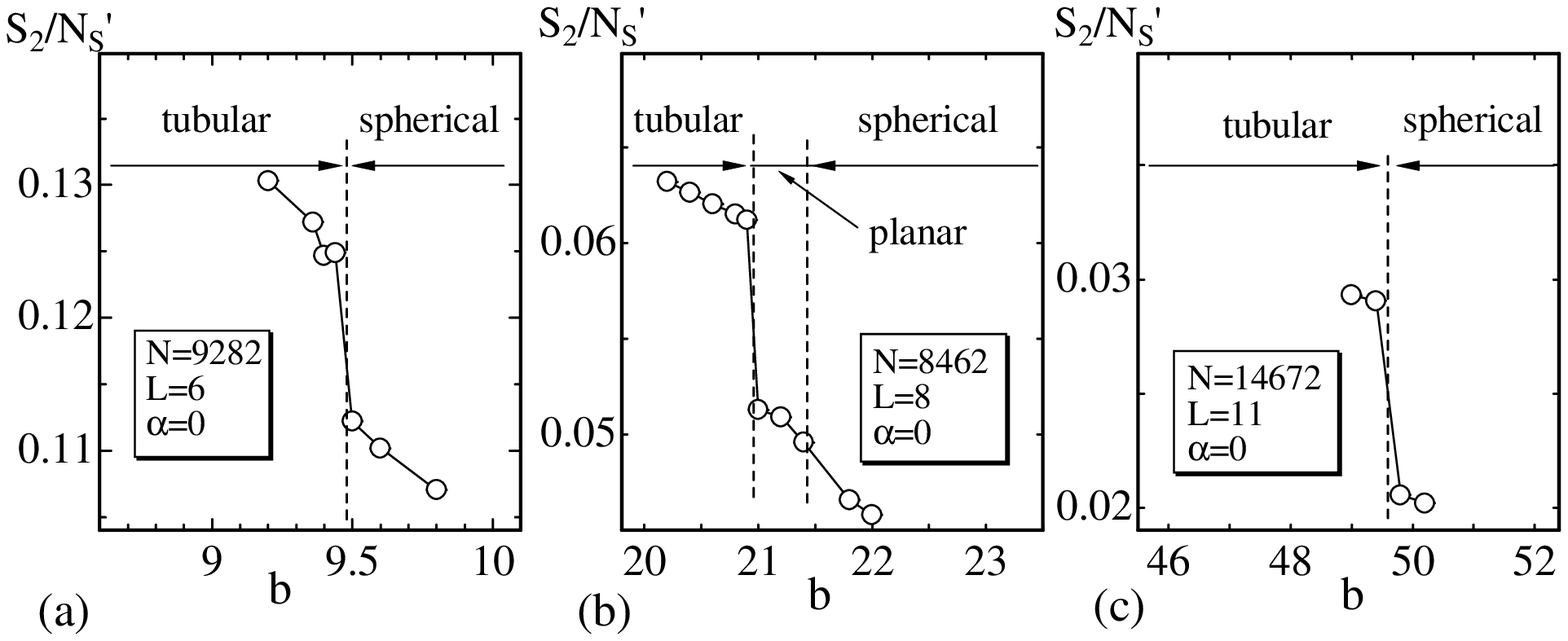}
\caption{The one-dimensional bending energy $S_2/N_S^\prime$ against $b$ obtained on the surface of (a) $L\!=\!6$, (b) $L\!=\!8$, and (c) $L\!=\!11$. $N_S^\prime(\!=\!N_S\!+\!6N_J\!-\!12)$ is the total number of vertices where $S_2$ is defined. } 
\label{fig-9}
\end{figure}
The one-dimensional bending energy $S_2/N_S^\prime$ obtained under $\alpha\!=\!0$ is shown in Figs.\ref{fig-9}(a)--\ref{fig-9}(c). A discontinuous change can be seen in $S_2/N_S^\prime$ not only in Fig.\ref{fig-9}(b) at the phase boundary between the tubular phase and the planar phase but also in Fig.\ref{fig-9}(c) at the phase boundary between the tubular phase and the spherical phase. A jump of $S_2/N_S^\prime$ in Fig.\ref{fig-9}(b) at the transition point between the planar phase and the spherical phase is very small, and hence is hardly seen just the same as in Fig.\ref{fig-3}(b) under $\alpha\!=\!3/2$ in the previous subsection. Thus, we find no difference between $S_2/N_S^\prime$ of $\alpha\!=\!0$ and that of $\alpha\!=\!3/2$.

\begin{figure}[htb]
\centering
\includegraphics[width=8.5cm]{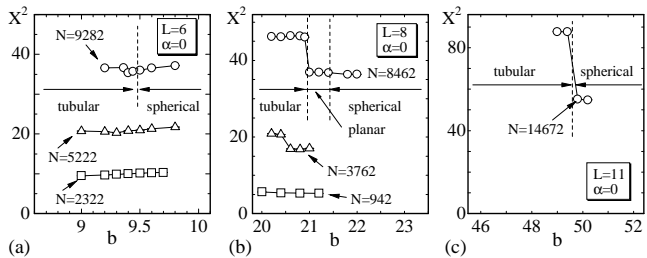}
\caption{The mean square size $X^2$ against $b$ obtained on the surfaces of (a) $L\!=\!6$, (b) $L\!=\!8$, and (c) $L\!=\!11$.} 
\label{fig-10}
\end{figure}
The mean square size $X^2$ are shown in Figs.\ref{fig-10}(a)--\ref{fig-10}(c). A jump is also seen in $X^2$ on the $L\!=\!8$ and $L\!=\!11$ surfaces in Figs.\ref{fig-10}(b) and \ref{fig-10}(c), and it is hardly seen on the $L\!=\!6$ surfaces of size up to $(N,N_S,N_J,L)\!=\!(9282,2400,162,6)$. These results are identical to those observed in  Figs.\ref{fig-7}(a)--\ref{fig-7}(c) under $\alpha\!=\!3/2$.

Finally, we comment on the planar phase appeared only on the $L\!=\!8$ surface. The thermal fluctuation of vertices inside the compartments disorders the surface against the bending energy of the compartment boundary. Therefore, the strength to disorder the surface increases (decreases) with increasing (decreasing) $L$ if $N$ remains fixed, as stated in Section \ref{model}. On the other hand, the mechanical strength of the surface increases (decreases) with decreasing (increasing) $L$, because the total number of junctions increases (decreases) with decreasing (increasing) $L$. Therefore, the strength to order the surface increases (decreases) with decreasing (increasing) $L$. Then, we expect that the surface is ordered (disordered) at sufficiently small (large) $L$ at given intermediate value of $b$. Moreover, it seems possible that two competitive forces to order/disorder the surface are balanced with each other at intermediate values of $L$ and consequently, some new phase appears depending on $b$ at those $L$. Note also that the possibility of the appearance of planar phase is not completely eliminated on the surfaces of $L\!=\!6$ and $L\!=\!11$ of sufficiently large size.

\section{Summary and conclusion}\label{Conclusions} 
We have shown that a dynamically triangulated spherical surface has three distinct phases;  the tubular phase, the planar phase, and the spherical phase, and that they are separated by discontinuous transitions. The first-order nature was very clear from the discontinuity in the bending energies $S_2$ and $S_3$ not only at the transition point between the tubular phase and the planar phase but also at the transition point between the tubular phase and the spherical phase. We know that the model has the collapsed phase at sufficiently small $b$, since the self-avoiding property is not assumed at least. Therefore, we expect that the model has four different phases including the collapsed phase, although the order of the transition between the collapsed phase and the tubular phase is unknown. 

 The mechanical strength of the surface is given only by elastic linear-chains with rigid junctions. The triangulated surfaces are characterized by the size $(N,N_S,N_J,L)$, where $N$ is the total number of vertices including the junctions, $N_S$ is the total number of vertices on the chains, $N_J$ is the total number of junctions, and $L$ is the length of chains between the two nearest-neighbor junctions on the starting configurations. These four parameters are not totally independent, because these are given by two independent integers $(\ell, m)$, where $m$ divides $\ell$. In fact, $N\!=\!10\ell^2\!-\!60m^2\!+\!2$, $N_S\!=\!30m(\ell\!-\!3m)$,  $N_J\!=\!10m^2\!+\!2$, and  $L\!=\!(\ell /m)\!-\!2$.

 We assumed three different values for $L$ such that $L\!=\!6$, $L\!=\!8$, and $L\!=\!11$ in the simulations. The edge length $R$ of the rigid junction was fixed to be $R\!=\!0.1$. The parameter $\alpha$, which represents a weight for the three-dimensional integrations of the partition function, was assumed as $\alpha\!=\!3/2$ and $\alpha\!=\!0$. 

It is remarkable that the model has the planar phase, which is stable only on the surfaces with a specific structure. In fact, the planar phase can be seen on the surfaces of $L\!=\!8$, and it is unstable on the $L\!=\!6$ and $L\!=\!11$ surfaces. The planar phase appears in a narrow region on the $b$-axis between the tubular phase and the spherical phase, and it is distinguishable from the spherical phase because a small but finite discontinuity can be seen in the bending energies $S_2/N_S^\prime$ and $S_3/N_B$. The gap of the bending energy $S_2$ at the transition point is very small, i.e., $S_2$ in the planar phase is almost identical to that in the spherical phase; however, the double peak structure was clearly seen in the histogram of $S_2$, which is included in the Hamiltonian. From this, we confirmed that the transition between the planar phase and the spherical phase is of first order. Our model in this paper indicates that one possible origin of planar shape of spherical membranes comes from the inhomogeneity due to the cytoskeltal structure and the fluidity of lateral diffusion of vertices.

We have confirmed that the results obtained at $\alpha\!=\!3/2$ in Eq.(\ref{Measure}) remain unchanged when $\alpha\!=\!0$. The phase structure of the fluid surface model in this paper is independent of the choice of $\alpha$ at least for $\alpha\!=\!3/2$ and $\alpha\!=\!0$. Large scale simulations should be performed. It remains to be studied how large $(\ell,m)$ are sufficient for the thermodynamic limit of the model.

\begin{acknowledgments}
This work is supported in part by a Grant-in-Aid for Scientific Research from Japan Society for the Promotion of Science.  
\end{acknowledgments}



\end{document}